\documentclass[12pt]{article}

\def\bea{\begin{eqnarray}}
\def\eea{\end{eqnarray}}

\def\ssc{\scriptscriptstyle}
\def\lsim{\mathrel{\raise.3ex\hbox{$<$\kern-.75em\lower1ex\hbox{$\sim$}}} }
\def\gsim{\mathrel{\raise.3ex\hbox{$>$\kern-.75em\lower1ex\hbox{$\sim$}}} }

\begin{document}
%\preprint{{\vbox{\hbox{NCU-HEP-k088}
%\hbox{Jan 2022}
%}}}
%\vspace*{.7in}

%\begin{frontmatter}
\begin{center}
%\title
{\large\bf Towards Noncommutative Quantum Reality\footnote{NCU-HEP-k088;
to be published in Stud. Hist. Phil. Sci. (2022)}
\\[.3in]
Otto C. W. Kong}\\[.2in]
{\bf Department of Physics and \\Center for High Energy and High Field Physics,\\[.1in]
%Center for Mathematics and Theoretical Physics, 
National Central University, Chung-li, Taiwan 32054}
\vspace*{.4in}
\end{center}
%\author{Otto C. W. Kong}
%\ead{otto@phy.ncu.edu.tw}

%\address{ Department of Physics and Center for High Energy and High Field Physics,
%Center for Mathematics and Theoretical Physics, 
%National Central University, Chung-li, Taiwan 32054  \\
%}

%\cortext[cor1]{Corresponding author.}

\begin{abstract}
\vspace*{.2in}
The implications of the physical theory 
of quantum mechanics on the question of realism is much 
a subject of sustaining interest, while the background 
questions among physicists on how to think about all 
the theoretical notion and `interpretation' of the theory 
remains controversial. Through a careful analysis of the 
theoretical notions with the help of modern mathematical 
perspectives, we give here a picture of quantum mechanics, 
as the basic theory for `nonrelativistic' particle dynamics,
that can be seen as being as much about the physical reality 
as classical mechanics itself. The key is to fully embrace the
noncommutativity of the theory and see it as a notion about
the reality of physical quantities. Quantum reality is then 
just a noncommutative version of the classical reality. 

\end{abstract}

Keywords : Quantum Mechanics, Quantum Reality
%\maketitle
\newpage
\section{Introduction}
Quine (1960) stated that ``What reality is like is the 
business of scientists". Among various disciplines of sciences, 
fundamental theories in physics seem to have a privileged 
position in informing us about the `basic essence of' reality.  
Quantum mechanics (QM) is the quantum version of the 
Newtonian theory of particle dynamics which fails to describe 
the `reality' of the atomic scale and beyond as successful as the 
former. However, almost a century after Heisenberg first put 
up the theory, what is the true meaning of that `quantumness', 
or what  `quantum reality' is like, remains controversial 
among physicists. Its conceptual foundations is still an area 
of active research. On the whole, what the basic situation 
seems to be is well illustrated from the following quotes 
from the introduction chapter of a book by Christopher 
Norris (2000) on the subject matter:\\
\indent{\em
 ``the orthodox `Copenhagen' interpretation of QM has 
influenced current anti-realist or ontologicalrelativist
approaches to philosophy of science" }\\
\indent{\em
``there are clear signs that some philosophers
 -- including Hilary Putnam -- have retreated from a realist 
position very largely in response to just these problems 
with the interpretation of quantum mechanics"}
\\
\indent{\em
``any alternative (realist) construal should have been so often 
and routinely ruled out as a matter of orthodox QM wisdom"}
\\
For a more comprehensive picture including more recent
perspectives, readers may consult Lombardi {\em et al.} (2019)
and  French and Saatsi  (2020). It suffices to say that the 
importance of the question sure maintains. In the latter book, 
for example,  French and Saatsi stated right from the beginning,  
(for ``quantum physics") that 
\\
\indent{\em
``what kind of knowledge does it provide us? This question gains 
significance from weighty epistemological issues that forcefully
arise in this context -- issues that are also at the heart of a more 
general debate on `scientific realism' in the philosophy of science."}

It is not however our intent to take a definite stand for scientific 
realism here. Our thesis is that \\
\indent{\em There is a way to look at quantum mechanics which 
makes it as much about a description of the physicists' reality as 
the classical, Newtonian theory, hence as intuitive or in
line with common sense as the latter.}\\
Hence, we want to argue against the ``notion of quantum mechanics 
as having destroyed the case for scientific realism" (Norris, 2000),
through discussions on a notion of {\em noncommutative quantum
reality} which we will argue to be a sensible natural way to look at 
the theory. Note that we are referring to classical physics broadly 
as Newtonian. To the extent that a background philosophical
perspective is relevant, we have to take no more than one of 
methodological naturalism. It should be emphasized that our
discussions are mostly focused only on the so-called `nonrelativistic'
particle theory, on which we have a solid mathematical description
of a notion of noncommutative values of physical quantities.  
To the extent that our argument stands, a parallel story for the 
`relativistic' and even field theory counterparts is in principle not 
difficult to think of, though establishing solid details of the 
mathematics and physics picture will take more effort. More on
the philosophical side, however, the picture of reality offered by 
quantum field theory in itself is a quite nontrivial topic, especially 
as the theory has been presented only with a picture of perturbation 
about free or noninteracting states.  As individual particles may be 
created or annihilated in the process, no set of particles can be taken 
as the entities the theory describes. Taking the theory seriously 
as a field theory with the fields as a sort of entities `living' in 
spacetime, as the commonly adopted picture of a classical field 
theory is also far from satisfactory. Quantum fields are formulated 
as operators which may not be Hermitian, hence may not even be
 taken as physical observables. So, they are neither observables nor 
states. Yet, all quantum fields in the theory are involved in any state,
at least as quantum fluctuations.
The states, in our opinion, can only be seen as states of the 
spacetime possessing different amount of the conserved quantities 
in different configurations. But no solid picture of such a model of 
spacetime has been given. Note that other than the electromagnetic
field, no other quantum field actually has a corresponding classical 
field theory. And the ontological nature of the electromagnetic field 
is far from unchallenged either (Lazarovici, 2018). Nevertheless, 
whatever physical or ontological picture one can otherwise give 
to the states of the theory, the notion of noncommutative reality 
can be applied to the ontology of the states and their observables 
as in quantum mechanics.

A word of caution to the readers is in order. 
The article mostly adopts a philosophical tone 
which may be seen to have taken some implicit assumptions
about what we want a physical theory to be, or at least how
we generally see the classical theory as. These include the 
mathematical entities and formalism of a physical theory
represent parts of physical reality, with the basic conceptually 
issues as intuitive or compatible with the scientifically based
intuition about nature (without devoting to any particular
mathematical theoretical models of such concepts). We see
achieving all that as the basic goal of theoretical physics, though 
past development on `interpretation' of quantum mechanics has 
`forced' some physicists to give up part of that. We make no 
apology for taking such a stand. However, readers may certainly
take a different stand there, or at least see all that as up to
debate and justifications. We emphasize again that we {\em only} 
present here our case for quantum mechanics being as  much about 
the physical reality and as intuitive as classical mechanics, and 
hope that readers will not bother too much about our philosophical 
tone on the classical theory being essentially intuitive and truly 
describing reality.

{\em Mathematics is the language of physical science}, which
gives rigorous logic of a theory and the required precision 
in its practical applications and hence verification. While
the conceptual foundation or interpretation of the theory
of quantum mechanics has been in `crisis', the theory is 
otherwise mathematically well defined. The debate on
how to think about the conceptual picture of the theory
keeps going on and on. To look into such conceptual issues,
a careful analysis about the basic mathematical notions
and the possible logically sounded conceptual thinking
about them, plausibly, and arguably most probably, beyond
the old conceptual framework is called for. Readers of
the current article are more likely to be quite familiar
with most of such mathematical notions in the theory of
quantum mechanics, and its classical counterpart, than 
otherwise. We will however address some of those very
carefully to help set the stage right for our discussions 
on the `quantum reality', and in so doing hopefully
makes the article more accessible to a broader readership.

It is important to note, however, that the theory of
quantum mechanics as a dynamical theory provides no 
dynamical description for the process of measurement.
The `standard' picture gives only a postulate about what 
is considered a successful measurement of a `quantum 
observable' on a particular `state' of the physical system 
being measured, namely one that gives a real number,
an eigenvalue, as the result in apparently the same way 
as a measurement of a physical quantity in classical, 
Newtonian, mechanics does. The postulate says the 
`state' will `collapse into' an eigenstate of the `observable'
the corresponding eigenvalue being that result obtained. 
The kind of measurement processes are called von
Neumann, or projective, measurements. The von Neumann 
measurements are far from the ideal kind, which are 
assumed not to change a state, and begs a dynamical 
theory for the successful description of the process 
itself. Achieving that within the basic framework of 
quantum mechanics is the task the decoherence 
theory (Zurek, 2003, 2018) sets for itself. 

In the famous EPR paper (Einstein, Podolsky, \& Rosen, 1935),
it is stated that\\
\indent{\em
``In a complete theory there is an element corresponding to 
each element of reality. A sufficient condition for the
reality of a physical quantity is the possibility of 
predicting it with certainty, without disturbing the system."}
\\
{\em Predicting} a physical quantity with certainty is
not the same as being able to measure or obtain its
value with certainty, and the authors were wise with their 
choice of words. While not being able to obtained the
value within a required uncertainty limit even in principle
sure renders `predicting it with certainty' meaningless,
not being able to do that with a single measurement of
a particular kind is a very different story. That is
a point we want our readers to bear in mind. 
Obsessions with a direct real number reading from an
apparatus as what nature tells us about the value of 
an observable are sure to be counter-productive. Hardly 
any precision measurement in physics nowadays shows
a simple output of the kind the understanding of which
is trivial and transparent. The upshot is as a single 
von Neumann measurement is far from all one can do to 
extract information about a physical quantity (Wiseman 
\& Milburn, 2010), how to think about the `reality' of 
the kind of measurement\footnote{
%%%%%%%%%%%%%% footnote#
The decoherence theory (Zurek, 2003,  2018) shows
that when a quantum particle is put to interact with 
an experimental apparatus for the measurement of
an observable to result in the apparatus settling into 
a `pointer state', {\em i.e.} giving a reading essentially 
like what we would have in a classical physics 
measurement, corresponding to an eigenvalue
of the observable, the state of the particle, the 
apparatus, and the environment around it generally
evolves quantum mechanically into some entangled
(micro)states. An entangled state of a system, 
nonexisting in classical physics, is one for which the 
notion of the state for a subsystem, here the particle, 
becomes ill-defined (Horodecki, 
Horodecki, Horodecki, \& Horodecki, 2009). 
The exact microstate at any moment cannot be
obtained theoretically or experimentally in any
practical sense due to the very large number of 
degrees of freedom beyond our control.  The 
picture hence becomes one of (quantum) statistical 
mechanics, where one knows only the macrostate
as described by the `pointer state'. The best one can 
do for describing the `state of the particle' is to get to
the reduced state, which is a mixed state as a statistical 
distribution. The theory shows the postulated von 
Neumann distribution could be obtained, without 
assuming the Born probability interpretation for 
the initial state of the particle measured. This is 
really a dynamical picture of the wavefunction
collapse. The particular eigenstate corresponding to 
the eigenvalue measured is more about the final state
of the particle after interacting with the apparatus
and the environment than the initial state one sets 
out to measure. Modern experiments [see for a
specially illustrative example Minev {\em et.al.} (2019)]
have essentially verified the collapse picture as
compatible with the decoherence theory.
}
%%%%%%%%%%%%%%%% end footnote
is actually a question of secondary importance to the key 
question about what `reality' is described by quantum 
mechanics as a physical theory. The so-called Heisenberg
uncertainty principle is often seen as giving a theoretical
uncertainty limit to the value of the physical quantities
obtained from the theory. We will illustrate that is
not quite true. {\em Quantum mechanics as a theory gives 
exact predictions about all physical quantities without 
any theoretical uncertainty, though there is difficulty 
for describing the values of such physical quantities in 
the same way as we do in classical mechanics. Hence,
we suggest the proper description of such values to be
done in the proper way, as the noncommutative values. }

Lastly, since we are going to rely much on the mathematical
notions in the formulations of quantum mechanics to be
the logical background for the conceptualization of the
{\em noncommutative quantum reality}, a few words on 
{\em mathematical reality} are in order. In line with
assuming only methodological naturalism, we further adopt 
{\em mathematical fictionalism}. We accept, for example, 
Mary Leng's (2010) statement ``reasons to believe in 
mathematical objects" as being real ``do not come from 
pure mathematics in empirical applications". Mathematical
objects, including the real numbers, are only abstract 
symbols useful for book keeping and logic reasoning, 
in relation to our scientific efforts to appreciate and
manipulate phenomena in Nature.  

Starting from the next section, we analyze the proper
notion of particle, space, and time, and then states and 
observables, leading to the central notion of the value 
of an observable for a state, for quantum mechanics from 
those of classical mechanics as the background. The key 
modern mathematical perspective of algebraic geometry
is the idea of a duality between an (observable) algebra 
and a geometric space (of its states). A state having fixed 
values for all observables is to be seen as giving an 
evaluation map of the full observable algebra. For 
a noncommutative algebra, like the algebra of quantum
observables, that suggests a notion of noncommutative
geometry beyond the picture of real number geometries.
We look at how such an evaluation map can be best 
formulated for the quantum observable algebra to fully
realize the duality, first to its space of pure states as the
quantum phase space, which leads to the introduction and
discussion about the notion of noncommutative values
of the physical quantities in section 3. The section 
finishes with the noncommutative coordinate,
hence noncommutative geometric, picture of the 
quantum space and the identification of it as the 
proper model for the physical space itself.
Some further discussions and conclusions
are given in the last section.

\section{Particle Dynamics Classical to Quantum}
\subsection{Particle, Space and Time}
In the standard physics presentations,  a Newtonian particle 
is a point mass  --- it is an object characterized by a single 
attribute or basic properties called {\em mass} which 
{\em occupies a single point in space}. The notion of space, 
or our physical space, and a position in it, is an intuitive one. 
So is time. To give the physical theory of particle 
dynamics, one needs to `quantify' position and time. 
As Quine put it, {\em ``To be is to be the value of a variable"}. 
We have in classical mechanics the position variable 
{\boldmath $x$} and time variable $t$. In Newton's time, 
there was essentially only one mathematical model of the 
physical space, that of a three-dimensional Euclidean geometry,
which he adopted. Hence, at every possible point in the 
physical space, the position is given by three real number 
values of its three coordinate variables $x_i$ ($i=1,2,3$), 
and time is taken to have the value of a single real number. 

If there is only one mathematical model for the physical space,
there is hardly a notion of right or wrong about it as a model.
In fact, one would take it as more than a model. It is then not 
difficult to appreciate the great philosopher Kant's rejecting the 
conception that space and time in Newtonian mechanics as physical 
entities, whether in his early Leibnizian ``relationalist" period or the
latter transcendental idealism. The best physical way to look at 
the {\em space} from the Newtonian\footnote
%%%%%%%%%%%%%
{Note that we used the term `Newtonian' throughout the article 
in the way it has been used in the modern physics literature, as 
essentially equivalent to `classical nonrelativistic'.}
%%%%
 theory, and arguably from 
any theory of particle dynamics, is that it {\em is the totality of 
all possible positions for a (free) particle}. Beyond that, the kind 
of physical theories cannot say anything more about the physical 
space, within its own logic. If there are different choice of 
available mathematical models, our logic here then would imply 
that the particular model for the physical space adopted by, or 
rather obtained from, a physical theory may only be as good as 
the theory itself. {\em When the Newtonian theory of classical 
mechanics is to be superseded, there is no good reason to assume 
it may be good to keep the Newtonian model of the physical space.}

Einstein's theory of special relativity replaces the Newtonian
space and time by a $1+3$ dimensional pseudo-Euclidean, or
Minkowski, spacetime. His theory of general relativity further
changes that to a  $1+3$ dimensional Riemannian geometry 
with a dynamical curvature. Apart from the Minkowski zero 
curvature limit, it is generally non-Euclidean. Quantum mechanics, 
however, keeps the classical notion of particle, space and time 
more or less untouched, except that there is a problem. 

The problem, or difficulty, quantum mechanics has with 
the Newtonian notion of particle, space and time is that 
a quantum particle cannot be required to always have 
a definite value for its single position in space. It is 
at least so stated almost in any literature on the subject
matter. However, there is a caveat. The statement should 
be more exactly stated as {\em a quantum particle cannot 
be required to always have a definite value for its single 
position in the Newtonian model of the physical space}.
{\em Whether the problem can be overcome in an 
alternative model for the physical space becomes then
a legitimate question and the theory itself may point 
at the right candidate for such a quantum model of the 
physical space.} The latter seems like a simple logical
possibility that has not been considered much in the 
literature on the physics of quantum mechanics or its 
background philosophy. 

\subsection{States and Observables}
A state of the classical particle is given by the six real 
number values of its position and momentum variables ({$x_i,p_i$}). 
The collection of all possible states forms the {\em phase space}, 
which is the six-dimensional Euclidean geometry with the
$x_i$ and $p_i$ as Cartesian coordinates. The six coordinates
are the basic observables, or physical quantities.

{\em Classical mechanics is a theory of real number valued
observables each of which is represented by a real variable.} 
Such variables, for the single particle, can be seen 
as functions of the six basic variables $x_i$ and $p_i$. 
When the six real numbers are known, the theoretical value 
of any one of its observables as a known function of the six 
basic coordinate variables $f(x_i,p_i)$ is determined in 
principle, without uncertainty. We have an algebra of 
observables as the collection of all observables, for which 
the sum of two of its elements, the product of a real number 
and an element, as well as the product of any two elements are 
all its elements. The algebra of observables for a classical 
particle is the usual algebra of functions $f(x_i,p_i)$.\footnote{
%%%%%%%%%%%%%% footnote
It is more convenient, and for the case of quantum mechanics
in a way necessary, to consider complex linear combinations of
observables in the algebra. In fact, it is generally accepted
that observable algebras for physical theories should be
$C^*$-algebras (Emch, 2009; Strocchi, 2008)
}
%%%%%%%%%%%%%%%% end footnote

A physical quantity in the quantum theory, the quantum 
observable, is described by a different kind of  mathematical 
object, the physical meaning of which is the crucial question 
we are after. The implications of all that about `quantum 
reality' actually hinges on the `actual values' of the 
observables for a particular state. For the quantum theory, 
whether a theoretical notion which is not an `observable' 
is anything physical is much the core of the sustaining 
controversial debate. The key focus is the question 
if the quantum state is observable, and whether it 
represents any `realistic' properties of the particle. 
A very recent article by Gao (2020) gives a nice summary 
about the debate on `ontic versus epistemic' state pictures 
or models. We are adding support to that `ontic' stand, from 
the conceptual point of view, by presenting a new picture on 
the relation between the states and the observables 
deemed impossible in much of the literature. We see
that relation as not so different conceptually for quantum 
mechanics as it is for classical mechanics, for which there 
is never such a debate. It should be noted that, from the
practical side, the modern quantum trajectory theory 
(Brun, 2002; Carmichael, 1993; Jacobs \& Steck, 2006;
Plenio \& Knight, 1998; Wiseman, 1996)  traces
the evolution of a quantum state through its deterministic
Schr\"odinger dynamics and probabilistic collapse well
and the theory has been very successfully applied to 
experimental settings. In particular, quite precise real-time 
monitoring of such trajectories in superconducting qubits
(Minev {\em et. al.}, 2019) verified a `realistic' picture
of the quantum state compatible with the decoherence 
theory. If a state of a physical system, in a `nonrelativistic' 
theory,  is a specification of its properties at an instant in time, 
a quantum/classical state should then be considered `real' 
if it gives the complete mathematical description of the 
mathematical content of the observables. Particle states in 
Newtonian mechanics sure has no problem there. We will see 
that the quantum particle state actually is also real in the same 
sense, that its mathematical description uniquely specified the 
mathematical content of all the observables and such
prediction can be experimentally verified. 

The geometric structure of the phase space,
its {\em symplectic geometry},  encodes all possible
dynamics for the particle. The mathematical structure 
can be taken simply as admitting Hamiltonian 
dynamics. Symplectic geometric structure can be used 
to formulate all classical dynamical systems and beyond, 
including special and general relativity, Maxwell's theory 
of electromagnetism, and actually quantum mechanics,
 as in the Hamiltonian formulations.
  
The phase space for a particle in the theory of quantum
mechanics is what is known as the projective Hilbert space
(Bengtsson \& \.Zyczkowski, 2006). Each ray of the 
infinite dimensional Hilbert space corresponds to a point
in the projective Hilbert space. The latter is a symplectic
geometry of infinite (real) dimension.
The picture is really better clarified only starting from 
about the 1980's. Very good description of that is
available from Cirelli, Mani\`a, and Pizzocchero (1990)
[see also Schilling (1996) and Roberts (2014)], within which 
there are very important further results we will get back to 
below. Details about Hamiltonian formulation of dynamics 
and their symplectic geometric picture is not necessary, in 
order to follow the present discussions. It suffices to note 
that the time evolution of the quantum state of a particle 
as given by the Schr\"odinger equation is exactly equivalent 
to an infinite pairs of differential equations, one for each
canonical pair of `position' and 'momentum' variables, all 
in the same mathematical form as those for the three pairs 
of $x_i$-$p_i$ variables for classical mechanics. All the 
`position' and 'momentum' variables, denoted by $(q_n,s_n)$ 
($n$ takes any positive integer value), are real variables, 
that is each of them takes the value of a real number on 
a fixed state. Looking at the phase space, quantum mechanics 
is not so different from classical mechanics at all except 
that {\em the theory really suggests the position of a quantum 
particle to be given by an infinite number of real numbers 
instead of only three.}

We introduced the state of a quantum particle as a point
on the specific geometric space instead of something to be
described by the Schr\"odinger wavefunction $\psi(x_i)$.
The notion of the wavefunction is a source of much confusion. 
Its form as a complex function of some $x_i$ variables, hence 
like a function on the Newtonian space, reinforces the idea of 
the theory being a theory on the Newtonian model of the 
physical space. It is also the background for the Born probability
interpretation of the orthodox `Copenhagen' picture, which gives 
up the intuitive notion that a particle has a definite position at all. 
The mathematics gives $\psi(x_i)$  as an infinite collection of 
complex number valued coordinates for the quantum 
phase space, but not that of the Born interpretation.  
If Schr\"odinger himself did start thinking about his 
formulation of quantum mechanics with the wavefunction 
$\psi(x_i)$ being a function on the physical space, he 
remained standing with Einstein against the orthodox 
`Copenhagen' picture of the theory.

The physical variables, or observables, of the theory is the
part that has little controversy. While the collection of all 
classical observables form a commutative algebra, that of 
the quantum observables form a noncommutative algebra. 
There is otherwise essentially a one-to-one correspondence 
between a classical and a quantum observable, at least so long 
as the standard usage of the terms in physics is concerned. 
The basic quantum observables are $\hat{X}_i$ and $\hat{P}_i$, 
which serve as the quantum position and momentum 
(coordinate) observables and the exact counterparts of the 
classical $x_i$ and $p_i$, respectively. They satisfy the 
Heisenberg commutation relation
\[
[\hat{X}_i, \hat{P}_j] \equiv \hat{X}_i \hat{P}_j - \hat{P}_j \hat{X}_i = i\hbar \,\delta_{ij}
\]
where $\delta_{ij}$ is unity for $i=j$ and zero otherwise, and 
$\hbar$ the Planck's constant divided by $2\pi$. We can still 
think about all other observables as elements of the 
noncommutative quantum algebra of observables to be given 
by a formal `function' of the basic observables, $f(\hat{X}_i, \hat{P}_i)$. 
One may naively project that $\hat{X}_i$ and $\hat{P}_i$ are like 
coordinates of the quantum phase space. Our question is like 
to what extent and how one can possibly make sense out of 
such a picture.  A more concrete mathematical picture takes the 
variables as linear operators on the Hilbert space. 
%In fact, there is a definite procedure 
%of {\em deformation quantization} (Hirshfeld and  Henselder 
%[2002]) for obtaining the quantum observable algebra from 
%the classical observable algebra. One simply introduces
 %a new noncommutative product between the observables 
%as like a kind of small modification of the simple, 
%commutative, classical product. A more concrete 
%mathematical picture takes the variables as linear 
%operators on the Hilbert space, including the 
%Schr\"odinger version of $\hat{X}_i$ given by $x_i$ and  
%$\hat{P}_i$ given by $-i\hbar\frac{\partial}{\partial x_i}$
%which has to be taken together, as differential operators 
%on the Hilbert space of his wavefunctions. 

If a quantum observable is not given as a real valued function 
of the basic variables, it certainly is not supposed to have 
a value of a single real number as the full information of the
observable for a particular state. Mathematically, the question 
is what can be taken as the value, the full information, 
of an operator for a particular state. We will argue that 
it should be a {\em noncommutative value}. {\em The 
noncommutative phase space coordinated by $\hat{X}_i$ 
and $\hat{P}_i$ is then exactly the space of all possible
(noncommutative) values of the coordinate variables,
a noncommutative geometry.}  Note that the Heisenberg 
uncertainty principle is really about the lack of precision 
in using single real numbers to model the values of the 
quantum observables, hence not at all in conflict with
the latter having definite noncommutative values.
The picture would agree perfectly with the theory if that
phase space can actually be identified with the projective 
Hilbert space mathematically. What amount effectively
to a transformation of coordinates between the infinite
number of complex number coordinates $z_n=q_n+i s_n$
and the three pairs of noncommutative coordinates 
$\hat{X}_i$ and $\hat{P}_i$ has been presented in
Kong and Liu (2021b), achieving that identification. Our
notion of noncommutative values for coordinates is not 
otherwise available in the literature of noncommutative 
geometry.  In particular, Huggett, Lizzi, and Menon (2021) 
illustrates the ``invalidity of the notion of localizability or a point"
 for noncommutative geometry as the ``undefinability" of 
``arbitrarily small separations", however, with smallness as to 
be characterized by a real number, exactly demonstrating the 
incompatibility of the notion of a real value for observables, 
coordinates or distance between points, with the geometry. 
The noncommutative values of physical quantities gives an 
alternative notion of `noncommutative point' which offers 
a coordinate picture for noncommutative geometry. 

\subsection{A Quantum State Can be Determined by Measurements.}
A quantum state can actually be represented by an 
observable. It is called the projection operator onto the 
state, or a density matrix for a `pure state'. So, a quantum 
state is essentially an observable. Can that observable really 
be observed? We actually have a direct positive answer to 
that from the domain of quantum optics, where physicists 
have been performing such measurements since the 1990's 
(Smithey, Beck, Raymer, \& Faridani, 1993; Leonhardt,
 1997, see also Minev {\em et.al.}, 2019). It is 
unfortunate that the line of works seems to be unknown 
to many theorists questioning the reality of the quantum 
state. Actually, that is about measuring the state of 
light in terms of what is called the Wigner distribution 
which is a representation of the projection operator for 
the state as a function of six real variables that look 
like those of the classical phase space variables $x_i$ 
and $p_i$. Measuring a function is certainly a lot more 
than measuring a number, but not anything experimental 
physicists are unfamiliar with. Mathematically, a function 
is like a set of infinite number of real numbers. Practically, 
for any finite precision approximate determination of it, 
obtaining a finite number of real number values for the 
function suffices. Recall that the standard geometric 
picture for the states is that each of them is a point in 
an infinite dimensional symplectic manifold, which can
be described by the infinite number of values for its real 
number coordinates. 

As a quantum state can be determined experimentally, 
through its projection operator or otherwise, up to any 
required finite precision at least in principle, it is of course 
as real as a classical state. Other important theoretical 
efforts to help `establishing' the `reality' of the quantum 
state include analysis based on the protective measurements 
(Gao, 2017, 2020), as well as the `ontic' versus
`epistemic' state analysis starting from the inspiring
paper of  Pusey, Barrett, and Rudolph (2012). Interested
readers may also consult Hardy (2013) and the references 
therein. We are illustrating here what we consider an 
intuitive and logically consistent and complete conceptual 
framework to look at the reality of quantum mechanics.

\subsection{On The Value of an Observable for a State}
%*********#
%The Born probability postulate was born in the midst of the physicists 
%trying to figure out what is the real number value a quantum observable 
%may be having
It is of crucial importance to carefully consider the role of 
a state in the notion of any kind of value for an observable. 
The latter is a variable because its value is not determined 
until we specify the state the system is in. Some philosophers 
and physicists seem to have forgotten that in their discussions 
about quantum reality, or the lack of. {\em If there is no 
reality in the state, the notion of the values for physical 
observables as part of reality is at best an ambiguous one.} 
Mathematicians indeed define a state of an algebra as 
a functional, let us call it an {\em evaluation functional}. 
{\em An evaluation functional is a map that sends each element 
of the algebra to a real number}. For the case of the observables
algebra for classical mechanics, that real number is the usual 
value the state has for the particular observable. {\em Each state 
can be identified exactly as one such evaluation functional.}  
More explicitly, we can use $[x_i,p_i]$ to denote the evaluation 
functional of a generic state, with the understanding that 
a definite state is to be given by definite real number values 
of the $x_i$ and $p_i$ in $[x_i,p_i]$. Then we can write
\[
[x_i,p_i](f) = f(x_i,p_i) \;,
\]
where for a particular state $[x_i,p_i]$, any observable
$f$ [or as the function $f(x_i,p_i)$ with  $x_i$ and $p_i$
as variables] is evaluated to have the real number answer 
of $f(x_i,p_i)$ at the specific values of $x_i$ and $p_i$.
This is the famous Gel'fand transform, which is the backbone 
of modern algebraic geometry and noncommutative geometry 
(Connes, 1994), on which we will come back to below.

One can start with the quantum observable algebra, and
define `states' as functionals of it. There is though a
mismatch of exact terminology one has to be careful of. 
The physicists' state as we have mostly been using the 
term above actually corresponds to a special class of 
functionals, called extremal states in mathematics or 
pure states in statistical physics. They send unity as an 
element of the algebra to the real number 1, and cannot 
be written as a nontrivial convex linear combination 
of other functionals (Alfsen \& Shultz, 2001). The 
mathematicians' generic state is used only in statistical 
physics with the name mixed states. Each of them describes 
a statistical distribution of pure states. Actually, with 
von Neumann as one of its pioneer, theory of operator 
algebras has found its applications in statistical quantum 
mechanics and quantum field theory (Brattelli \& Robinson,
 1987; Emch, 2009; Swanson, 2020) since not too long after 
the  publication of his book on the mathematical foundation 
of quantum mechanics (von Neumann, 1955). The pure and 
mixed state terminology has been getting more and more 
popular in physics (Bengtsson \& \.Zyczkowski, 2006), 
though most discussions about fundamental theories still 
used the term state without the word pure, as we do above. 
In fact, as we do not have to care much about the mixed 
states beyond this paragraph, we will mostly simply keep
using the term state for a pure state below, when the 
context leads to no confusion. What is important to 
note is that the collection of mixed states is just the 
collection of convex linear combinations of pure states. 
Statistical states of course can be constructed as such
from the single particle states.

The mathematical notion of (pure) states for the quantum
observable algebra agrees exactly with the physical states
of the theory. Each point $(q_n,s_n)$ of the projective Hilbert 
space can be seen as an evaluation functional that maps every 
observable given by the operator $\hat{A}$ to the definite real 
number value $[q_n,s_n](\hat{A})$. The number is the expectation 
value of the observable for the state. Practically, it is the 
average of the statistical distribution of eigenvalue results 
from von Neumann measurements. From the theoretical point of 
view, the expectation value is certainly the best candidate 
for a single real number value of a quantum observable, and 
is definitely predicted by the theory. Moreover, it can be
experimentally determined up to any required precision, at 
least from von Neumann measurements. A direct measurement 
of it can also be implemented as a protective measurement 
(Aharonov, Anandan, \& Vaidman, 1993, 1996). 

Quantum mechanics has definite notions of the physical
quantities as a noncommutative observable algebra and
the states as evaluation functions giving the definite
expectation values. That is what the mathematical logic 
of the theory presents. Why is that not good enough?
There has been the prejudice of taking the single
eigenvalue result from a von Neumann measurement 
as the true value, which is tied with the Born probability 
interpretation of the quantum state, adhering to taking 
the Newtonian model for the physical space. More 
importantly, the picture cannot be easily connected to 
its classical approximation and the classical notion 
of the position and momentum observables. We can 
consider all possible expectation values of the six 
$\hat{X}_i$ and $\hat{P}_i$. The collection does not give 
a picture of the phase space. The best one can do is to 
associate each set of the six values to a coherent state
 (Peremolov, 1977). One can easily see that different 
states can have exactly the same set of the expectation 
values. This is an indication that the expectation value 
does not carry the full information of the observable on 
the state, which can also be easily appreciated from
the statistical distribution it is associated with. 
Actually, the full distribution is definitely predicted 
by the theory, and it is possible to be determined up
to any required precision experimentally, at least 
in principle. Such a distribution certainly carries
more information about than only its expectation value.
And if two distributions measured for two states share
the same expectation values but are otherwise very
different, we sure do not want to say they share the
same value for the observable. To describe that 
difference, we need more than the expectation value. 
Maybe we can think about taking as the value the
full distribution, say by noting the infinite set of
moments of the distribution. The noncommutative value
is something in the direction, only better.

\section{The Noncommutative Values of Physical Quantities and the Noncommutative Space}
\subsection{The Quest for the Noncommutative Value and the Evaluation Homomorphism}
We have brought up a couple of limitations of the notion
of states as evaluation functionals in quantum mechanics.
While each state as a point in the phase space corresponds
to such a functional for each observable, their real 
number values for the position and momentum observables 
are not enough to fully distinguish the state. Such a  
value does not give full information about the observable
for the state, for which one may need an infinite number of 
real numbers. A careful thinking would reveal a further
deficiency of the evaluation functionals for the quantum 
observable algebra compared to the classical one when one 
looks at the relation between such functionals. The classical 
state as an evaluation map is really a homomorphism; that 
is to say that it maps each observable as a real variable 
onto a real number value in such a way that the whole 
observable algebra is mapped into a subalgebra of the 
algebra of real numbers with the algebraic relations among 
the variables preserved by their values. Mathematically, 
we have the variable $fg$ as a product of variables $f$ 
and $g$ as an example,
\[
[x_i,p_i](fg) = fg(x_i,p_i)
= f(x_i,p_i)\,g(x_i,p_i)=[x_i,p_i](f)\, [x_i,p_i](g)\;.
\]
$fg(x_i,p_i)$ as a function equals to the product of 
functions $f(x_i,p_i)$ and $g(x_i,p_i)$ by definition, 
and that simply applies to the values of the functions 
at any point. In fact, a relation of this kind is very 
important in physics, any relationship between the 
observables predicted theoretically can only be 
verified by checking their values. The evaluation map 
for the quantum observables taken as given by the 
evaluation functionals giving the expectation values 
obviously fails to maintain that. In fact, no 
functional can do that. Functionals have real numbers 
as values, which commute among themselves. But the 
quantum observables do not. If one can find a 
functional $[\omega]$ that, for example, satisfies
\[
[\omega](\hat{X}_i \hat{P}_i) = [\omega](\hat{X}_i)\, [\omega](\hat{P}_i)\,,
\]
to which the expectation value functional certainly fails,
we would have
\[
[\omega](\hat{P}_i) \,[\omega](\hat{X}_i) = [\omega](\hat{X}_i)\, [\omega](\hat{P}_i) \;,
\]
giving the result 
\[
[\omega](\hat{X}_i \hat{P}_i) = [\omega](\hat{P}_i \hat{X}_i)
\]
for the operator products $\hat{X}_i \hat{P}_i$ and
$\hat{P}_i \hat{X}_i$ evaluated on any state $[\omega]$.
The latter is as good as the statement that the
observables $\hat{X}_i \hat{P}_i$ and $\hat{P}_i \hat{X}_i$
are really the same. However, the basic structure of
quantum observable algebra is exactly the nontrivial
commutation relation.

Following the above, we can see that it would be a good
idea to generalize the notion of an evaluation functional
with a real number value to a noncommutative analog 
with a value that is an element of a noncommutative
algebra keeping the evaluation map as a homomorphism
between the quantum observable algebra and the 
algebra of its values. Moreover, we would like each 
such noncommutative value to have the information 
content of infinite number of real numbers covering all 
information in the probability/statistical distribution 
obtainable from von Neumann measurements,
with the expectation value playing an important role
inside. One can even think about the algebra of all 
such noncommutative values as like the set of
 `{\em noncommutative numbers}'. It is of interest 
to note that the preface of one of Takesaki's books 
(2003) on the theory of operator algebra starts 
with the sentence
\\
\indent {\em
``The author believes that the theory of operator 
algebras should be viewed as a number theory in analysis."}
\\
We see that with the notion of the noncommutative values,
the operators, or quantum observables, can be seen rather 
as noncommutative number variables. Together with the
noncommutative value notion, a complete picture of
`noncommutative number theory' may evolve. 

Furthermore, the six $\hat{X}_i$ and $\hat{P}_i$ observables 
to serve as a system of noncommutative coordinates of the
quantum phase space as the projective Hilbert space, we 
need again each of the noncommutative coordinates to have 
a value carrying the information of an infinite number of 
real numbers. One can think about the value of a quantum 
observable as a piece of quantum information
(Braunstein \& van  Loock, 2005), mathematical seen as
an element of a noncommutative algebra, each of which can 
be seen as like an infinite number of classical information
({\em i.e. real numbers});
\\
\indent {\em ``a single qubit can 
substitute for an infinite number of classical bits"}, 
\\
as Hardy (2013) put it (Galv\~ao and Hardy, 2003).

\subsection{A Solid Answer to Our Call}
Though the notion of the noncommutative value has not
been introduced by other authors, the solution candidate 
has been available. For each quantum state, a one-to-one
homomorphism, between observables in quantum 
mechanics and a noncommutative algebra of so-called 
`symmetry data' with each element as a set of infinite 
number of real numbers, has been given by Schilling (1996),
in a Ph.D dissertation under the supervision of Ashtekar.
We want to skip much of the technical details here only 
to sketch the essence of the story. {\em We emphasize
that each noncommutative value is one mathematical
quantity}, an element of a noncommutative algebra
much like a real number is an element of a commutative 
algebra. {\em It is not a necessity to think about it as
the set of infinite number of real numbers}, though
a convenience before we learn to deal with the
noncommutative values directly. 

The Ashtekar-Schilling homomorphism is a map for each state
that takes a quantum observable $\hat{A}$ as an operator 
essentially to the set of values of all the derivatives 
of expectation value function $[\omega](\hat{A})$ taken 
as a function of the state $\omega$, say, explicitly as 
functions of the real coordinates $(q_n,s_n)$. Let us put
it as $[\omega](\hat{A})= f_{\!\ssc\hat{A}}(q_n,s_n)$. 
By all derivatives here, we include the zeroth derivative,
which is just the expectation value function itself. 
Actually, only the derivatives up to the second order are 
needed as the higher order derivatives for the class of 
functions can be expressed in terms of the zeroth, first, 
and second order ones, and the known metric of phase
space. With infinite number of coordinates ($n$ counting 
from 1 to $\infty$), we have an infinite set of real number 
values. The derivatives of the expectation value function 
for the operator product $f_{\!\ssc\hat{A}\hat{B}}(q_n,s_n)$ 
can be given in terms of  the derivatives of 
$f_{\!\ssc\hat{A}}(q_n,s_n)$ and 
$f_{\!\ssc\hat{B}}(q_n,s_n)$ that verifies the
evaluation map we put as $\{[\omega]\}$ sending
an observable $\hat{A}$ to $\{[\omega]\}(\hat{A})$,
the latter being the noncommutative value of `symmetry
data', as a homomorphism between the observable 
algebra $\hat{A}$ and the noncommutative algebra 
of $\{[\omega]\}(\hat{A})$. For example, with the
noncommutative, but associative, product as the 
rules to give the $f_{\!\ssc\hat{A}\hat{B}}(q_n,s_n)$ 
derivatives marked by $\star_{\kappa}$, we have
\[
\{[\omega]\}(\hat{A}\hat{B}) = \{[\omega]\}(\hat{A}) \star_{\kappa} \{[\omega]\}(\hat{B}) \;,
\]
and 
\[
\{[\omega]\}(\hat{A}\hat{B}) - \{[\omega]\}(\hat{B}\hat{A})
 = \{[\omega]\}(\hat{A}) \star_{\kappa}\{[\omega]\}(\hat{B}) -  \{[\omega]\}(\hat{B}) \star_{\kappa}\{[\omega]\}(\hat{A})\;,
\]
for each $\{[\omega]\}$ on any observables $\hat{A}$
and $\hat{B}$. The map is of course a linear one. The
noncommutative value to be identified as the set of
infinite number of real numbers should each be seen
directly as what it is, namely an element in a
noncommutative algebra.

%%%%%%
\[
{\{{\lbrack\omega\rbrack\}}{(\widehat A\widehat B)}}={\{{\lbrack\omega\rbrack\}}{(\widehat A)}}\star_\kappa{\{{\lbrack\omega\rbrack\}}{(\widehat B)}},
\]
%%%%%%%%

Explicitly, more conveniently expressed in terms of derivatives with 
respect to the complex coordinates $z_n=q_n+i s_n$ taking value at
$z_n = \tilde{z}_n=\tilde{q}_n+i \tilde{s}_n$ characterizing
a particular $\{[\omega]\}$, we have a description of
the noncommutative value as 
 $\{[\omega]\}(\hat{A}) = \{ f_{\!\ssc\hat{A}},
 \frac{\partial  f_{\!\ssc\hat{A}}}{\partial z_1},
\frac{\partial  f_{\!\ssc\hat{A}}}{\partial z_2}, \cdots , 
\frac{\partial  f_{\!\ssc\hat{A}}}{\partial z_1 \partial \bar{z}_1},
\frac{\partial  f_{\!\ssc\hat{A}}}{\partial z_1 \partial \bar{z}_2},\cdots , 
 \frac{\partial  f_{\!\ssc\hat{A}}}{\partial z_2 \partial \bar{z}_1}, \cdots 
\}$,  where  $\bar{z}_n=q_n-i s_n$ and all functions 
inside the expression are to be evaluated at 
$z_n = \tilde{z}_n=\tilde{q}_n+i \tilde{s}_n$. Only 
up to second order derivatives are needed. Getting
the noncommutative product $\{[\omega]\}(\hat{A}\hat{B})$
from $\{[\omega]\}(\hat{A})$ and  $\{[\omega]\}(\hat{B})$
is just above calculating the expressions of the (complex
number values) of $f_{\!\ssc\hat{A}\hat{B}}$, 
 $\frac{\partial  f_{\!\ssc\hat{A}\hat{B}}}{\partial z_n}$ and
$\frac{\partial  f_{\!\ssc\hat{A}\hat{B}}}{\partial z_m \partial \bar{z}_n}$,
for $m,n=1$ to $\infty$, from those of $f_{\!\ssc\hat{A}}$
and $f_{\!\ssc\hat{B}}$ following the noncommutative
product $\star_{\kappa}$ as essentially explicitly given
by Cirelli, Mani\`a, and Pizzocchero (1990). The complex
number values in the sequence of complex coordinate 
derivatives can certainly be expressed in term of real
number values for the real coordinates $q_n$ and $s_n$,
only that the expression is formally more complicated.
Conceptually, we mostly talk about them here in terms of
the real ones. From the real number values of $f_{\!\ssc\hat{A}}(q_n,s_n)$ 
and $f_{\!\ssc\hat{B}}(q_n,s_n)$ at a point, one cannot
get the corresponding value for $f_{\!\ssc\hat{A}\hat{B}}(q_n,s_n)$.
However, from the values of all the derivatives of 
 $f_{\!\ssc\hat{A}}(q_n,s_n)$ and $f_{\!\ssc\hat{B}}(q_n,s_n)$,
one can retrieve from the noncommutative product $\star_{\kappa}$ 
the values of all the derivatives of $f_{\!\ssc\hat{A}\hat{B}}(q_n,s_n)$.
That infinite set of values for the derivatives is what is
to be seen as a single noncommutative value.

\subsection{The Noncommutative Geometric Picture}
Although Ashtekar and Schilling were apparently not aware
of the important work of Cirelli, Mani\`a, and Pizzocchero 
(1990) when performing their study, quite a part of
their results have been given in the latter paper with,
in a way, a more direct picture. For example, Ashtekar 
and Schilling mostly worked with commutators and
anticommutators instead of the products. The latter
authors have also given an interesting isomorphism, which
can be seen as a logical precursor of the Ashtekar-Schilling 
map. The Cirelli-Mani\`a-Pizzocchero isomorphism
involves essentially a noncommutative associative product,
called the K\"ahler product, between the expectation value 
functions $f_{\!\ssc\hat{A}}(q_n,s_n)$, which we write as
\[
f_{\!\ssc\hat{A}\hat{B}}(q_n,s_n)
 = f_{\!\ssc\hat{A}}(q_n,s_n) \star_{k} f_{\!\ssc\hat{B}}(q_n,s_n)
\]
or equivalently
\[
[\omega](\hat{A}\hat{B}) = [\omega](\hat{A}) \star_{k} [\omega](\hat{B}) \;.
\]
The product involves differentiation with respect to the
coordinates. Having the product between the functions of
course implies relations between the derivatives and hence
their values.  While the quantum observables as operators
naively cannot be seen as functions $f(q_n,s_n)$, the 
expectation value functions $f_{\!\ssc\hat{A}}(q_n,s_n)$ 
can be written as $f_{\!\ssc\hat{A}}(q_n,s_n) \star_{k} 1$,
hence each corresponds to a differential operator
$f_{\!\ssc\hat{A}}(q_n,s_n) \star_{k}$ involving the 
derivatives of $f_{\!\ssc\hat{A}}(q_n,s_n)$. The algebra
of operators $f_{\!\ssc\hat{A}}(q_n,s_n)\star_{k}$ is 
isomorphic to the observable algebra of $\hat{A}$. In 
fact, they should really be seen as the same algebra 
described differently. With the noncommutative K\"ahler
product, the quantum observable algebra may then be
seen as an algebra of functions on the quantum phase
space as its space of pure states. That fulfills the basic idea 
of noncommutative geometry, except that the actual 
geometry does not look, naively, any noncommutative.

To see that quantum phase space as a noncommutative
geometry, recall that $\hat{A}=A(\hat{X}_i,\hat{P}_i)$.
Then, the alternative description of $\hat{A}$ as
$f_{\!\ssc\hat{A}}(q_n,s_n)$ is exactly saying that
the six $\hat{X}_i$ and $\hat{P}_i$ may be seen as an
alternative set of coordinate variables in the place 
of the infinite set of $q_n$ and $s_n$. So, each point
of the quantum phase space can be specified by the
infinite real number values of the real coordinate
variables $q_n$ and $s_n$, or the six noncommutative
values of noncommutative coordinate variables 
$\hat{X}_i$ and $\hat{P}_i$. Another way to see 
that is to think about the infinite number of the
derivatives of the six $f_{\!\ssc\hat{X}_i}(q_n,s_n)$
and $f_{\!\ssc\hat{P}_i}(q_n,s_n)$ as the alternative
infinite number of real number coordinates. Physicists 
and mathematicians are very familiar with describing
the same geometry with different choices of coordinate
systems and making transformations among them, 
including having complex number valued coordinates
each of which has the information content of two real
number coordinates. Going between the commutative 
and noncommutative coordinates (Kong \& Liu, 2021b)
may simply be a new manifestation of the kind. 

Looking further into quantum dynamics from the 
perspective of the noncommutative coordinates,
the theory can be seen as giving equations of motion
for the time development of any observable, or rather
the noncommutative value of the observable. That is
the Heisenberg equation of motion. The equation can 
be `transformed' into a form given in terms of any
of the time dependent $f_{\!\ssc\hat{A}}(q_n,s_n)$ 
function for the evolving observable, and the 
$f_{\!\ssc\hat{H}}(q_n,s_n)$ for the physical 
Hamiltonian operator or energy observable. The 
resulted equation is the Poisson bracket form of 
equation of motion for a classical-like observable 
with the classical-like energy observable as the 
generator of time evolution in the usual Hamiltonian 
formulation (Johns, 2005). Hence in exactly the same 
mathematical form as for classical mechanics except 
with a different phase space. The generic equation 
applied to the functions $f_{q_n}(q_n,s_n)=q_n$ 
and $f_{s_n}(q_n,s_n)=s_n$ give essentially
the Schr\"odinger equation of motion. Moreover,
the Heisenberg equation of motion itself should 
be seen as exactly an equation of the kind with
$\frac{1}{i\hbar}$ times the operator commutator
seen as the Poisson bracket for the quantum
observables as $\hat{A}=A(\hat{X}_i,\hat{P}_i)$
(Cirelli, Mani\`a, \& Pizzocchero, 1990;
Schilling, 1996). Note that 
expression for the quantum operator, or 
noncommutative variable, now appreciated as the
quantum Poisson bracket, was identified early to give 
the classical Poisson bracket at the classical limit by Dirac. 
Again,  that simply suggests thinking about $\hat{X}_i$ 
and $\hat{P}_i$ as pairs of noncommutative canonical 
coordinates of the phase space, the position and 
momentum coordinate observables indeed for the 
phase space as a symplectic geometry. One can 
actually obtained from it the standard form of 
Hamilton's equation of motion for $ \hat{X}_i$ 
and $\hat{P}_i$ :
\[
\frac{d}{dt} \hat{X}_i = \frac{\partial \hat{H}}{\partial \hat{P}_i} \;, 
\qquad
\frac{d}{dt} \hat{P}_i = - \frac{\partial \hat{H}}{\partial \hat{X}_i} \;,
\]
where we have the physical Hamiltonian operator
$\hat{H}=H(\hat{X}_i,\hat{P}_i)$, to be matched to 
\[
\frac{d}{dt} {x}_i = \frac{\partial {H}(x_i,p_i)}{\partial {p}_i} \;, 
\qquad
\frac{d}{dt} {p}_i = - \frac{\partial {H}(x_i,p_i)}{\partial {x}_i} \;,
\]
in classical mechanics with ${H}(x_i,p_i)$ as the Hamiltonian 
function (Kong \& Liu, 2021b). From the mathematical studies, 
there are in fact good indications that noncommutative 
geometries, as the spaces of pure states, may all be in 
a way symplectic (Chen, 2014;
Cirelli, Lanzavecchia, \&  Mani\`a, 1983; 
Roberts \& Teh, 2016; Shultz, 1982).

Physicists are familiar with the transformation
between the Heisenberg picture and the Schr\"odinger
picture of quantum dynamics. The standard interpretation
of that is to depict the time evolution for the same
expectation value of an observable on a state (vector)
in terms of either the evolving observable in the
former or the evolving state in the latter. With
the Hamiltonian formulation perspective for both
the Heisenberg and the Schr\"odinger equations of
motion, they are simply about depicting the time
evolution for the observable, or its noncommutative 
value, in terms of the noncommutative coordinates 
as $A(\hat{X}_i,\hat{P}_i)$ or the commutative 
coordinates as $f_{\!\ssc\hat{A}}(q_n,s_n)$,
respectively. That is exactly a coordinate 
transformation picture.

We note in passing that the notion of the
operator `coordinate observables', or noncommutative
coordinate variables, $\hat{X}_i$ and $\hat{P}_i$ as
coordinates for the noncommutative geometry is very
much a pure physics one. It is based on quantum
mechanics as the quantum version of classical mechanics.
Mathematicians may otherwise not be particularly 
interested in the specific algebra. While the notion of 
local coordinates can be used to define a commutative 
geometry as a real manifold, which is locally (around 
any particular point) essentially a copy of a Euclidean 
geometry, noncommutative geometries are to be defined 
algebraically with a basic pure geometric picture under 
pursuit. What is a local picture of a noncommutative 
geometry? What is the noncommutative analog of an 
Euclidean space? One cannot even be sure that such 
questions are sensible ones. Physicists have however 
been working on noncommutative spacetime models 
essentially based on the idea of the noncommutative 
position, and time, coordinate variables (Doplicher, 
Fredenhagen, \& Roberts, 1995; Wess \& Zumino,
 1990; Wess, 2007), commonly believed to be 
necessary for Planck scale physics. To keep a good
control on the otherwise speculative nature of such 
physical theories, the nature of the physics for the
noncommutative coordinate variables of quantum 
mechanics may be a very useful guideline, 
especially if that also gives a noncommutative 
model of the physical space.

\subsection{The Quantum Physical Space}
We have painted almost a full picture of the 
noncommutative quantum reality above. Quantum 
mechanics is to be seen as a theory with physical quantities 
described by noncommutative observable variables, taking 
noncommutative values on a state. The state of a particle
is a point in the noncommutative symplectic geometry of 
its quantum phase space, to be fully characterized by the 
noncommutative values of the six noncommutative position 
and momentum coordinate observables. All of that is 
exactly like the case of classical mechanics if the
adjectives `noncommutative' are replaced by `commutative',
of course also `quantum' replaced by `classical'. The
classical theory is exactly the commutative approximation
to the quantum one. One thing is however missing. That is
the quantum or noncommutative picture of the physical space.

A naive answer to the question may be the physical space 
as seen from quantum mechanics is the totality of all 
possible noncommutative values for the three position 
observables $\hat{X}_i$. Not only that we do not have 
a commutative geometric picture about it, the idea has
quite some problems. Considering the three 
$\hat{X}_i$ and `functions' of them only, there is zero
noncommutativity. And it is not clear what may be the
role of the algebra in physics,  On the other hand, there 
is actually a simple argument telling that the quantum 
phase space, unlike its classical counterpart, cannot be 
properly seen as a product of the space of positions and 
the space of momenta. It is so exactly in the same way 
the Minkowski spacetime as the spacetime model 
from Einstein's theory of special relativity cannot be
properly seen as a product of a three dimensional space 
and a one dimensional time. Technically, it is a question 
of the spacetime model as a representation space of the 
relevant (relativity) symmetry (Chew, Kong, \& Payne, 2017). 
Minkowski spacetime, seen as a vector space, corresponds 
to what is called an irreducible representation of the 
Lorentz symmetry. Its Newtonian approximation, with 
the Lorentz transformation symmetries replaced by 
the Galilean counterparts, is a reducible representation 
that reduces to Newtonian space and Newtonian time 
as irreducible components.\footnote{
%%%%%%%
The mathematical description of any system or object 
observing a certain set of symmetries is known to 
correspond essentially to a representation of the group
as the mathematical description of the set of symmetries.
A representation is reducible if it can be seen from the 
symmetry picture, as consisting of a few independent
parts. Otherwise, it is irreducible. Hence, to the extent that
space or spacetime is seen to have a certainly (relativity)
symmetry, its mathematical description should correspond
to such a representation (Elliott \& Dawber, 1979).
}%%%%%%%%%%%% 
~For the quantum mechanics, part of the 
relevant symmetry is the Heisenberg-Weyl symmetry 
which incorporates the noncommutativity between the 
$\hat{X}_i$  and $\hat{P}_i$,  the quantum Hilbert space 
is essentially the only representation space obtainable
from an irreducible representation (Taylor, 1986). Hence, 
the quantum phase space is the only sensible model for the 
quantum physical space. While within any specific Lorentz 
frame of reference the Minkowski spacetime can be seen
as a product of a Newtonian space and the geometric 
space/line of time, an admissible Lorentz boost to another 
frame of reference would take the time line to a new 
direction which is partly an original spatial direction. As
all the Lorentz frames are on equal footing,  no part or 
subspace, no line, in the Minkowski spacetime can be 
definitely identified as what time is all about. Hence the Minkowski 
spacetime has to be seen as a single `entity'. That is the
notion of the Minkowski spacetime as an irreducible 
representation. Quantum mechanics as an analog case has 
a fundamental symmetry which gives a complex phase 
transformation to the Hilbert space vectors actually 
without observable consequence. It takes the real 
number coordinates $q_n$ and $s_n$, to $q'_n$ and 
$s'_n$ with $q'_n+i s'_n=e^{i\theta}(q_n+i s_n)$. The 
configuration/position (sub)space of described then by the  
$q'_n$ coordinates is not the same as the one described by 
the $q_n$ coordinates. There is no reference frame 
independent way to identify a configuration/position subspace.

It is of interest to note that within the mathematics of 
noncommutative algebra, there is actually a notion of 
dimension (Mc Connell \& Robson, 2001) relevant to 
the quantum picture of the physical space here. The 
(projective) Hilbert space taken as a cyclic module 
of the noncommutative ring of polynomials in 
$\hat{X}_i$  and $\hat{P}_i$ has a (Gel'fand-Kirillov) 
dimension of three, which agrees with our intuitive 
notion of dimension for the physical space. In fact,
the corresponding cyclic module for the algebra generated
 by $n$ pairs of $X_i$-$P_i$ has Gel'fand-Kirillov dimension 
of exactly $n$. The three pairs of $X_i$-$P_i$ are certainly
(linear) independent,  though each pair has the fixed 
Heisenberg commutation relation.  That is at least how 
far we see that the notion of Gel'fand-Kirillov dimension 
matches the intuitive picture familiar in their classical, 
commutative, limit. In the latter, Newtonian, limit,
however, $x_i$ and $p_i$ become independent too
and the position space and momentum spaces 
exist as separable notions.

\section{Further Discussions and Conclusions}
Wess (2007) had remarked that
\\
\indent {\em ``That a change in the concept of space 
for very short distances might be necessary was 
already anticipated in 1854 by Riemann
in his famous inaugural lecture."}
\\
as he also quoted from Riemann (1854)
\\
\indent {\em ``Now it seems that the empirical notions 
on which the metric determinations of space are founded, 
the notion of a solid body and a light ray, cease to be 
valid for the infinitely small. We are therefore quite 
at liberty to suppose that the metric relations of space 
in the infinitely small do not conform to the hypotheses 
of geometry; and we ought in fact to suppose it, if we 
can thereby obtain a simpler explanation of phenomena
\dots} \\ 
\indent {\em \dots The answer to these questions can 
only be got by starting from the conception of phenomena 
which has hitherto been justified by experience, and 
which Newton assumed as a foundation, and by making 
in this conception the successive changes required by 
facts which it cannot explain. Researches starting from 
general notions, like the investigation we have just 
made, can only be useful in preventing this work from 
being hampered by too narrow views, and progress in 
knowledge of the interdependence of things from being 
checked by traditional prejudices."}
\\
He further noted that
\\
\indent {\em `` \dots from the discovery of quantum
mechanics. There physics data forced us to introduce 
the concept of noncommutativity."}
\\
and advocated the idea of noncommutative coordinates,
only not so much for the phase space or the physical 
space of quantum mechanics itself. Of course a consistent 
picture of the latter cannot be obtained without the 
notion of the noncommutative values. We echo, again, 
Quine's {\em ``To be is to be the value of a variable"} 
here. The {\em ``very short distances"} most physicists
working on noncommutative geometry have in mind 
are much shorter than the relevant scale of quantum 
mechanics, for which we already actually have {\em 
``the notion of a solid body and a light ray"} as 
classical physics knows them {\em ``cease to be valid"}. 
The picture of noncommutative quantum reality does
give a {\em ``simpler explanation of phenomena"} even 
in line with our intuition so long as that is not
{\em ``checked by traditional prejudices"}. 

Let us take up further on the {\em ``traditional prejudices"} 
of physical quantities having real number values. Bohr 
emphasized a lot about all measurements as giving classical, 
here to be read as real number, results. Upon more careful 
thinking, the real numbers we have been reading out of our 
measuring apparatus are always really from the real number 
scales we put onto the apparatus during our calibrations. Even
in reading a simple pointer position, we get the real number
answer only because we use the real number geometry model 
to look at the physical space in which the pointer and the scale 
sit. Otherwise, the real number value is never indicated by 
Nature. In line with {\em mathematical fictionalism} (Leng, 2010) 
is Quine's notion of {\em ``convenient fiction"}. For the case of 
quantum physics, however, real numbers as part of that old 
fiction is not even convenient. The noncommutative values are 
the new convenient fiction. Maybe we should simply deal with 
the noncommutative values directly as pieces of quantum 
information. Indeed, all kinds of quantum information 
experiments can be seen as giving some measurements on 
a quantum system. We are extracting information about the 
system under study, which can be used, together with our
theory, to get information about the initial state of the system. 
How one may directly deal with the noncommutative values 
or quantum information is a very nontrivial question which 
requires ingenious thinking in practical experimental settings, 
to which we are not capable of giving any specific suggestion 
at this point. The world is quantum hence all information
obtainable from it of quantum nature, though a major part of 
those we have learned to manipulate can be well approximated 
by classical, real number, information. It is more like we have
succeeded, with our classical physical theories, to deal with 
physical phenomena in which the model of real number valued 
physical quantities seems to offer a good enough description, as 
in classical physics measurements and projective measurements
for quantum physics, rather than the our world being so much
as apparently commutative. Of course technically the quantum 
theory gives conditions when its classical counterpart works well 
enough as an approximation. Our science and technology is only 
starting to deal with quantum information is relatively simple 
but yet extremely unclassical settings. There is a very long way 
to go. A more humble approach is to determine the infinite
sequence of complex numbers $\{[\omega]\} (\hat{A})$ as a 
description of a noncommutative value up to a certain precision 
with probably a good number of real number measurements, 
which would not impose much difficulty in principle. 
Otherwise, a lot more effort on the part of theoretical and
experimental physics may be needed for us to gain more
understanding of the noncommutative value picture.

It is of interested to note that in 1934, Haldane
(1934) spoke against quantum mechanics as a
 {\em ``refutation of materialism"}, calling it 
 {\em ``a refutation of \dots\dots spatialism"}.
Haldane's {\em spatialism} is  {\em ``the Cartesian 
view of matter as definitely localized in space"}.
Our noncommutative quantum reality picture presents
the theory as not a refutation of spatialism either, 
but rather more like only a refutation of the 
simple Cartesian view of space, namely the 
Newtonian space model itself. Haldane also
remarked about that {\em ``mental events are 
inexactly localized in space-time"}, where the
space-time of course means the Newtonian models
of them. A puzzling question is then how well
can we see mental events as localized in 
noncommutative quantum models of space-time.

The business of science is about describing `reality',
and fundamental physical theories try to do that
with conceptual and mathematical models. Going from
classical mechanics to quantum mechanics as a more
successful theory, we have changed our models for
the observables and the states. The mathematical
perspective of noncommutative geometry or algebraic 
geometry says that the observable algebra and the
phase space should be seen as dual descriptions of
one another. Hence, the `reality' is actually to be
described by the relations among the values of all
the observables, with the state being given by the
values of the basic observables, the position and 
momentum ones. To properly understand quantum
mechanics as the relations among the `real' 
values of the observables, we also have to change 
the classical model for such values, to the 
noncommutative one. The evaluation homomorphism 
then fully illustrates the duality. The model for the 
physical space, with `positions' or locations, points, 
to be described by the coordinate observables, has
also to be changed from the classical Newtonian 
model. Only then we have a fully consistent story 
for that quantum reality. 
{\em The simple move to fully embrace 
the quantum noncommutativity of Nature actually 
allows us to see quantum mechanics as being as much about 
the noncommutative reality as classical mechanics is 
about the assumed commutative reality.}\footnote{ 
  %%%%%%\footnote
While the article focuses on the philosophical presentation
of the notion of quantum reality as offered by a new
theoretical perspective on the theory of quantum mechanics  
taking the noncommutativity more seriously than ever as the
core feature and properties of Nature it depicts, the physics
and mathematics of the key new conceptual notion of the 
noncommutative values of physical quantities has been
studied in Kong (2020, 2021), Kong \& Liu  (2021a)
in the meantime.}\\

\noindent{\bf Acknowledgments}

The author is partially supported by research grant number 
109-2112-M-008-016 of the MOST of Taiwan.

\end{document}